\documentclass[aps,prd,reprint,superscriptaddress,floatfix]{revtex4-2}
\usepackage{amsmath,amssymb,graphicx,hyperref}
\usepackage[utf8]{inputenc}
\usepackage[T1]{fontenc}
\usepackage{xcolor}

\begin{document}

\title{The scheme independent $3$-sphere free energy is not a monotone $F$-function}
\author{Giacomo Santoni}
\email{giacomo.santoni@alumni.sns.it}
\affiliation{Physics Department, INFN Roma1, Piazzale A.\ Moro 2, Roma, I-00185, Italy}

\author{Francesco Scardino}
\email{francesco.scardino@uniroma1.it}
\affiliation{Physics Department, INFN Roma1, Piazzale A.\ Moro 2, Roma, I-00185, Italy}
\affiliation{Physics Department, Sapienza University, Piazzale A.\ Moro 2, Roma, I-00185, Italy}

\begin{abstract}
We study the natural scheme-independent quantity obtained from the three-sphere partition function of a $(2+1)$-dimensional quantum field theory by removing all local counterterm ambiguities. At conformal fixed points this quantity equals the standard $F$-theorem invariant. Conformal perturbation theory shows that it locally decreases at $O(g^2)$ under any relevant scalar deformation of a three-dimensional CFT. However, an exact analysis of the free massive scalar on $S^3$ shows that this sphere-free-energy interpolant is not monotone along the full renormalization-group flow: it dips below its infrared value and then returns to it. Thus the natural counterterm-subtracted quantity built from sphere thermodynamics is not, by itself, a monotone $F$-function. We trace the obstruction to the second-order differential operator required to eliminate the local ambiguities.
\end{abstract}

\maketitle
\section{Introduction} A central question in quantum field theory is which observables can encode RG irreversibility. In $(2{+}1)$ dimensions the $F$-theorem states that $F\equiv -\log|Z(S^3)|$, evaluated at conformal fixed points, satisfies
$F_{\rm UV}\ge F_{\rm IR}$~\cite{Jafferis:2011,Klebanov:2011,CasiniHuerta:2012}. The proof~\cite{CasiniHuerta:2012} constructs a monotone $F$-function from the
entanglement entropy of a disk, $\mathcal{F}_{\rm EE}(R)=S_{\rm EE}(R)-R\,S_{\rm EE}'(R)$, whose monotonicity follows from strong subadditivity (SSA). A long-standing and natural idea is that the sphere partition function itself could
provide a monotone interpolant, without invoking entanglement. Such a quantity would be directly accessible to lattice computation and would place
the $F$-theorem on the same thermodynamic footing as the $c$-theorem in $d=2$~\cite{Zamolodchikov:1986,Affleck:1986,BloteCardyNightingale:1986,Nakayama:2013is}.

\par The obstacle is that the sphere free energy $W_{S^3}(R)\equiv\log|Z(S^3_R)|$ is contaminated by scheme-dependent local counterterms that for a maximally symmetric metric scale as $R^3$ and $R$. In general, $Z(S^3_R)$ includes also a gravitational Chern-Simons counterterm, but in $W_{S^3}(R)$ it is removed by the modulus. Since such local terms also shift the first and second logarithmic derivatives of $W_{S^3}$,
no monotonicity statement for the bare sphere free energy is meaningful before removing these ambiguities.
\par In this letter we construct a ``double filter'' that removes these ambiguities and yields a scheme-independent $F$-function $F_{\cal E}(R)$, prove that it decreases perturbatively near any UV fixed point for every relevant deformation, and then show---by an exact free-field computation---that it fails to be monotone along the full RG flow. The failure is structural: the sphere partition function requires two UV subtractions, yielding a second-order differential filter with no analogue of SSA to control its sign. This settles negatively the idea that sphere thermodynamics alone can encode monotone RG irreversibility in odd dimensions so that any monotone $F$-function must use additional structure such as entanglement (SSA) or spectral positivity.

\section{Construction} On $S^3_R$ the only UV-sensitive local counterterms allowed by diffeomorphism invariance are~\cite{BirrellDavies:1982,Vassilevich:2003}
\begin{align}
W_{\rm loc}=\lambda_0\!\int\!\sqrt{g}+\lambda_1\!\int\!\sqrt{g}\,\mathcal{R}
\end{align}
We parametrize the surface of $S^3_R$ by three ``angles'' $\chi \in [0,\pi)$, $\theta \in [0, \pi)$, $\varphi \in [0, 2\pi)$. The metric tensor is
\begin{align}
    ds^2=R^2\left(d\chi^2+\sin^2\chi\left(d\theta^2+\sin^2\theta d\varphi^2\right)\right)
\end{align}
and the Ricci scalar is
\begin{align}
    \mathcal{R}=\frac{6}{R^2}
\end{align} 
It follows that
\begin{align}
    \int_{S^3_R}\sqrt{g}=2\pi^2R^3\ , \qquad \int_{S^3_R}\sqrt{g}\mathcal{R} = 12\pi^2R
\end{align}
We define the differential operator $D\equiv R\,\partial_R$ and the double filter
\begin{align}
    \label{eq:double_filter}
    \mathcal{D}_{\rm ren}^{(3)}\equiv
    \Big(1-D\Big)\!\Big(1-\tfrac{1}{3}D\Big)
    =1-\tfrac{4}{3}D+\tfrac{1}{3}D^2,
\end{align}
which is the unique polynomial of degree $\le 2$ in $D$ satisfying $\mathcal{D}_{\rm ren}^{(3)}[R^3]=\mathcal{D}_{\rm ren}^{(3)}[R]=0$ and $\mathcal{D}_{\rm ren}^{(3)}[1]=1$. The thermodynamic $F$-function is
\begin{align}
    \label{eq:FE_def}
    F_{\cal E}(R)\equiv -\mathcal{D}_{\rm ren}^{(3)}\,W_{S^3}(R).
\end{align}
By construction, $F_{\cal E}$ is invariant under $W\to W+\lambda_0 R^3+\lambda_1 R$
and reduces to $F=-\log|Z(S^3)|$ at conformal fixed points. The endpoint inequality $F_{\cal E}^{\rm UV}\ge F_{\cal E}^{\rm IR}$ follows from
the $F$-theorem~\cite{CasiniHuerta:2012}. As a differential operator acting on $W_{S^3}(R)$, the filter in Eq.~\eqref{eq:double_filter} can in principle be implemented on the lattice by computing $W$ at nearby radii (e.g.\ via thermodynamic integration) and using finite differences for $D W$ and $D^2 W$, with no model-dependent UV subtraction.

\section{Perturbative monotonicity} We first show that $F_{\cal E}$ decreases near the UV at leading nontrivial order in conformal perturbation theory (CPT). Consider a CFT on $S^3_R$ deformed by a relevant scalar $\mathcal{O}$ of dimension $\Delta<3$, with coupling $g$ of mass dimension $3-\Delta$.

\par Since $\langle\mathcal{O}\rangle_{S^3}=0$ for $\Delta>0$, the leading correction to $W$ is $O(g^2)$ and involves the integrated two-point function.
For a scalar primary, $\langle\mathcal{O}(x)\mathcal{O}(y)\rangle=C_{\mathcal{OO}}\big/\big(2R\sin\tfrac{\theta}{2}\big)^{2\Delta}$, where $\theta$ is the geodesic angle between $x$ and $y$.
Fixing one point on $S^3$ and using $d^3y\sqrt{g}=4\pi R^3\sin^2\!\theta\,d\theta$, one finds
\begin{align}
    \label{eq:twopt}
    \int_{S^3_R}\!\!\int_{S^3_R} d^3x\sqrt{g}\,d^3y\sqrt{g}\,\langle\mathcal{O}(x)\mathcal{O}(y)\rangle
    &= C_{\mathcal{OO}}\,\omega(\Delta)\,R^{6-2\Delta},\nonumber\\
    \omega(\Delta)&=16\pi^{7/2}2^{-2\Delta}\,\frac{\Gamma\!\big(\tfrac32-\Delta\big)}{\Gamma(3-\Delta)}.
\end{align}
The integral is absolutely convergent for $\Re\Delta<3/2$ and is defined for $\Delta\ge 3/2$ by analytic continuation after subtraction of local counterterms~\cite{Klebanov:2011,Pufu:2016}. For $\Delta\neq 3/2,\,5/2$, the coefficient of $R^{6-2\Delta}$ is scheme-independent, since finite local counterterms
on $S^3$ can only shift $W$ by terms proportional to $R^3$ and $R$. We recall that $C_{\mathcal{OO}}>0 $ by unitarity. Inspection of Eq. \eqref{eq:twopt} reveals that this integral has no scheme ambiguity unless $\Delta=3/2$ or $\Delta = 5/2$ because in those cases it scales as $R^3$ and $R$ respectively. Accordingly,
\begin{align}
    \label{eq:W_CPT}
    W_{\rm univ}(R)=-F_{\rm UV}+\tfrac12 g^2 C_{\mathcal{OO}}\,\omega(\Delta)\,R^{6-2\Delta}+O(g^3),
\end{align}

\par The filter removes the aforementioned scheme ambiguity by acting as $\mathcal{D}_{\rm ren}^{(3)} [R^{6-2\Delta}]=Q(\Delta)\,R^{6-2\Delta}$ with $Q(\Delta) =\frac{4}{3}(3/2-\Delta) (5/2-\Delta)$, which is negative for $3/2<\Delta<5/2$.
Naively this would produce $dF_{\cal E}/d\log R>0$ in this window. However, the renormalized coefficient $\omega(\Delta)$ also changes sign because $\Gamma(3/2-\Delta)<0$ for $\Delta>3/2$. The product simplifies via the Gamma recursion:
\begin{align}
    \label{eq:gamma_identity}
    Q(\Delta)\,\frac{\Gamma(\tfrac32-\Delta)}{\Gamma(3-\Delta)}
    =\frac{4}{3}\,
    \frac{\Gamma(\tfrac72-\Delta)}{\Gamma(3-\Delta)}\;>0
\end{align}
for all $0<\Delta<3$. Therefore 
\begin{align}\label{eq:deriv}
    \frac{dF_{\cal E}}{d\log R}\Bigg\rvert_{O(g^2)}=-\kappa(\Delta)\,\frac{1}{2}g^2 C_{\mathcal{OO}}\,R^{6-2\Delta}<0
\end{align}
with $\kappa(\Delta)\equiv \omega(\Delta)Q(\Delta)>0$, for every relevant deformation. The sign flip in the filter polynomial is exactly compensated by the sign flip in $\omega(\Delta)$ -- a structural consequence of the filter being built to annihilate precisely the counterterms whose subtraction controls the sign of $\omega(\Delta)$.

\section{Exact free-scalar computation} The CPT check probes $mR\ll 1$.
We now show that monotonicity fails nonperturbatively.

\par For a free conformally coupled scalar of mass $m$ on $S^3_R$, the eigenvalues of $-\nabla^2+\tfrac{3}{4R^2}+m^2$ are $\lambda_n=(n^2-\tfrac14)/R^2+m^2$ with degeneracy $n^2$~\cite{Klebanov:2011}. From $W=-\tfrac12\sum_{n=1}^\infty n^2\log\lambda_n$ one obtains $DW=\sum_{n=1}^\infty n^2(n^2-\tfrac14)/(n^2-\tfrac14+x^2)$ with $x\equiv mR$. Writing $n^2/(n^2+\alpha)=1-\alpha/(n^2+\alpha)$ with $\alpha\equiv x^2-\tfrac14$
and using $\zeta(-2)=0$, $\zeta(0)=-\tfrac12$~\cite{AbramowitzStegun:1965}, and the Mittag--Leffler identity (analytic continuation in $\alpha$ understood)~\cite{GradshteynRyzhik:2007}
$\sum_{n=1}^\infty 1/(n^2+\alpha) =\tfrac12(\pi\alpha^{-1/2}\coth(\pi\sqrt{\alpha})-\alpha^{-1})$, one obtains
\begin{align}
    \label{eq:DW_closed}
    DW(x)
    =\frac{\pi x^2}{2}\,\nu(x)\coth\!\big(\pi\nu(x)\big),
    \quad
    \nu\equiv\sqrt{x^2-\tfrac14}.
\end{align}
For $x<1/2$ analytic continuation gives $\nu\,\coth(\pi\nu)\to\sqrt{\tfrac14-x^2}\,\cot(\pi\sqrt{\tfrac14-x^2})$. Since $\mathcal{D}_{\rm ren}^{(3)}$ commutes with $D$,
the flow derivative $dF_{\cal E}/d\log R = -\mathcal{D}_{\rm ren}^{(3)}[DW]$
requires only derivatives of this known smooth function.\par 

In general differentiating a divergent series term by term and then applying zeta regularization is not a valid procedure: the operations do not commute. The counterexample $\sum_{n=0}^\infty(n+a)$ demonstrates this clearly.\par

We show in Appendix \ref{app:rigor} that the result in the paper is nevertheless
correct, by performing our calculation following two routes: regularize first, then differentiate and differentiate first, regularize then. The agreement is not a coincidence: it is guaranteed by a structural property of the spectral zeta function on $S^3_R$. That is, $Z(0,\alpha)=0$ independently of the mass parameter~$\alpha$. This property ensures that the potential discrepancy between the two orderings vanishes identically.\par 

We also point out that any possible discrepancy given by the regularization procedure, that doesn't break diffeomorphism invariance, would have been eliminated by the filter regardless, since scheme dependent terms can only arise precisely in the form that the filter is designed to remove.

\par
\begin{figure}[t]
    \centering
    \includegraphics[width=\columnwidth]{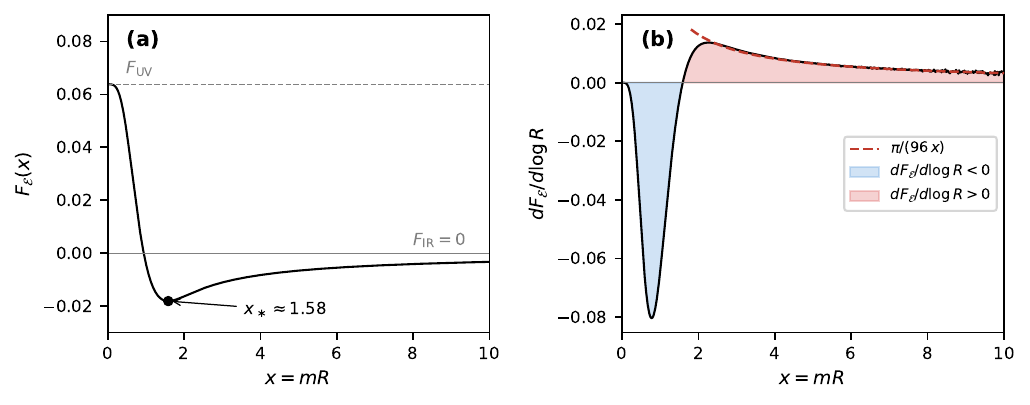}
    \caption{%
    (a) Thermodynamic $F$-function $F_{\cal E}(x)$ for a free massive scalar on $S^3$ ($x=mR$).
    $F_{\cal E}$ starts at $F_{\rm UV}\approx 0.064$ (dashed line), overshoots below
    $F_{\rm IR}=0$, reaching a minimum at $x_\ast\approx 1.58$ (dot), then returns to $0^-$ from below. (b) Flow derivative $dF_{\cal E}/d\log R$.
    Negative for $x<x_\ast$, positive for $x>x_\ast$ (shaded regions). The dashed curve on the top right corner is the analytic asymptotic $\pi/(96\,x)$ in Eq.~\eqref{eq:large_x}, which controls the large-$x$ tail. The sign change demonstrates that $F_{\cal E}$ is not a monotone $F$-function.
    \label{fig:FE}}
\end{figure}

 At large $x$, $DW\sim\frac\pi2 x^3-\frac{\pi}{16}x -\frac{\pi}{256}x^{-1} +\cdots$ so the filter kills $x^3$ and $x$. Therefore, since $\mathcal{D}_{\rm ren}^{(3)}[x^{-1}]=\tfrac{8}{3}\,x^{-1}$, the $-\tfrac{\pi}{256}x^{-1}$ term gives the leading survivor:
\begin{align}
    \label{eq:large_x}
    \frac{dF_{\cal E}}{d\log R}\xrightarrow{x\to\infty}
    \frac{\pi}{96\,x}+O(x^{-3})>0.
\end{align}
Since the derivative is negative in the small $x$ regime by the perturbative analysis, as shown in Eq.~\eqref{eq:deriv}, and positive for large $x$ as in Eq.~\eqref{eq:large_x}, it must change sign. A numerical evaluation (Fig.~\ref{fig:FE}) gives the zero crossing at $x_\ast\equiv mR_\ast\approx 1.584$. Integrating from $x=0$ where $F_{\cal E}=F_{\rm UV}\approx 0.0638$, the function reaches
\begin{align}
    F_{\cal E}(x_\ast)\approx -0.0181 \approx -0.28\,F_{\rm UV},
\end{align}
overshooting below $F_{\rm IR}=0$ before slowly returning as $F_{\cal E}(x)\to -\pi/(96\,x)\to 0^-$. 

\section{Structural explanation} The non-monotonicity is not an accident of the free theory, indeed, it reflects a general obstruction tied to the order of the differential filter.

\par The entanglement $F$-function $\mathcal{F}_{\rm EE}=S_{\rm EE}-R\,S_{\rm EE}'$ uses a \emph{first-order} filter, because the entanglement entropy of a disk has a single UV divergence (the area law $S_{\rm EE}\sim R/\epsilon$). Its derivative $d\mathcal{F}_{\rm EE}/d\log R=-R^2 S_{\rm EE}''$ involves only $S_{\rm EE}''$, whose sign is fixed by SSA~\cite{CasiniHuerta:2012}.

\par By contrast, $W_{S^3}$ has two UV divergences ($R^3$ and $R$), necessitating the second-order filter of Eq.~\eqref{eq:double_filter}. The derivative $dF_{\cal E}/d\log R$ now involves both $D^2 W$ and $D^3 W$, and no known positivity condition simultaneously constrains both. Equivalently, the filter polynomial $Q(\Delta) =\frac{4}{3}(3/2-\Delta) (5/2-\Delta)$ must change sign between its roots $\Delta=3/2$ and $\Delta=5/2$ where at $O(g^2)$ in CPT this is compensated by the renormalized coefficient in Eq.~\eqref{eq:gamma_identity}, but the exact computation shows that this compensation fails at $mR\sim O(1)$.

\par More generally, such polynomial filters of degree $n\ge 2$ in $D$ that annihilate $n$ independent power laws necessarily have sign changes in their filter polynomial, suggesting a generic obstruction to producing monotone interpolants from $W$ alone, absent additional dynamical input (analogous to SSA) that controls the higher-order variations of $W$.

\section{Discussion} The quantity $F_{\cal E}$ is not a monotone $F$-function, but it remains useful: it provides a UV-finite, scheme-independent diagnostic that correctly captures $F$ at fixed points, satisfies $F_{\rm UV}\ge F_{\rm IR}$, and can be implemented on the lattice via finite differences without model-dependent subtraction. However, practitioners using sphere free energies to track degrees of freedom along RG flows -- whether in numerical bootstrap, lattice simulations, or holographic comparisons -- should be aware that monotonicity is not guaranteed, even in free theories. The essential point is that removing all local scheme ambiguities is not sufficient to promote
the sphere partition function to a monotone measure of degrees of freedom.

\par The structural lesson is that the number of UV subtractions determines the order of the differential filter, and filters of order $\ge 2$ are generically sign-indefinite. Monotone $F$-functions require structure beyond counterterm removal: either a positivity condition on second variations such as SSA for entanglement entropy or a spectral representation like the dilaton effective action for the $a$-theorem~\cite{KomargodskiSchwimmer:2011}. The sphere partition function provides neither, and that is why thermodynamics alone cannot replace entanglement as a probe of RG irreversibility.

\par The broader replica-energy framework, including applications to topological phases,
fracton order, and the $d=4$ anomaly analogue, will be presented elsewhere~\cite{companion}.

\appendix
\section{Rigorous derivation of $D W$}
\label{app:rigor}

For a free conformally coupled scalar of mass $m$ on $S^3_R$, the relevant kinetic operator is $A=-\nabla^2+\tfrac{3}{4R^2}+m^2$. Its eigenvalues and degeneracies are
\begin{equation}\label{eq:eigenvalues}
  \lambda_n = \frac{n^2-\tfrac14}{R^2}+m^2,
  \qquad d_n = n^2,
  \qquad n=1,2,3,\dots
\end{equation}
We define $x\equiv mR$ and $\alpha\equiv x^2-\tfrac14$, so that
\begin{equation}
  \lambda_n = \frac{n^2+\alpha}{R^2}.
\end{equation}
The sphere free energy is $W=\log|Z|=-\tfrac12\log\det A$. Formally, the divergent sum is
\begin{equation}
\label{eq:wbare}
    W_{\rm bare} = -\frac{1}{2}\sum_{n=1}^{\infty}n^2\log \lambda_n = - \frac{1}{2}\sum_{n=1}^{\infty}n^2\log\frac{n^2+\alpha}{R^2}
\end{equation}

\subsection{Regularization vs differentiation}
We will now perform the two key steps in two orders
\begin{enumerate}
    \item[(a)] Route 1: Regularization $\Rightarrow$ differentiation
    \item[(b)] Route 2: Differentiation $\Rightarrow$ regularization
\end{enumerate}
For future use we introduce the function
\begin{equation}\label{eq:Zdef}
    Z(s,\alpha)=\sum_{n=1}^\infty n^2(n^2+\alpha)^{-s}
\end{equation}
that converges absolutely for $s>3/2$ and extends meromorphically to all $s\in\mathbb{C}$. Using $n^2 =(n^2+\alpha)-\alpha$ we can express $Z(s,\alpha)$ as
\begin{align}\label{eq:Zdecomp}
    Z(s,\alpha)=H(s-1,\alpha) -\alpha H(s,\alpha)
\end{align}
where $H(s,\alpha)$ is
\begin{equation}\label{eq:Hdef}
    H(s,\alpha)=\sum_{n=1}^\infty (n^2+\alpha)^{-s}
\end{equation}
These functions satisfy the shift identities
\begin{align}\label{eq:shifts}
    &\partial_{\alpha}H(s,\alpha)=-sH(s+1,\alpha), \nonumber \\
    &\partial_{\alpha}Z(s,\alpha)=-sZ(s+1,\alpha).
\end{align}

Since $Z(s,\alpha)$ is jointly analytic in $s$ and $\alpha$ in a neighbourhood of $s=0$ (for $\alpha\notin\{-n^2\}_{n\geq 1}$), the mixed partial derivatives commute:
\begin{equation}\label{eq:mixed}
  \partial_\alpha \partial_s Z(s,\alpha)\rvert_{s=0} = \partial_s\!\left[\partial_\alpha Z(s,\alpha)\right]\Big|_{s=0}.
\end{equation}
Also, another property that will be used below is \cite{Kirsten:2001wz}
\begin{align}\label{eq:zero}
    &Z(0,\alpha) = 0 && \forall \alpha > -1
\end{align}

\paragraph{Regularization $\Rightarrow$ differentiation} 
The renormalized free energy is then
\begin{align}\label{eq:Wreg}
    W = & \frac12\partial_{s}[R^{2s}Z(s,\alpha)]\rvert_{s=0} \nonumber \\
    =&\log R\ Z(0,\alpha) +\frac12Z_s(0,\alpha),
\end{align}
where $Z_s(0,\alpha) \equiv \partial_sZ_s(s,\alpha)\rvert_{s=0}$. Thanks to \eqref{eq:zero} this simplifies to,
\begin{equation}
  W = \frac12Z_s(0,\alpha)\ .
\end{equation}



We need $DW=R\,\partial_R\, W$. Since $\alpha=m^2R^2-\tfrac14$ and $R\,\partial_R\alpha=2m^2R^2=2x^2$ (at fixed $m$), we have
\begin{equation}
\label{eq:DW_route1}
  DW = x^2\,\partial_\alpha Z_s(0,\alpha).
\end{equation}
Using~\eqref{eq:mixed} and the shift identity~\eqref{eq:shifts}:
\begin{align}
  \partial_\alpha Z_s(0,\alpha)
  &= \partial_s\!\left[\partial_\alpha Z(s,\alpha)\right]\Big|_{s=0}
  = \partial_s\!\left[-s\,Z(s{+}1,\alpha)\right]\Big|_{s=0} \nonumber\\
  &= \Big[-Z(s{+}1,\alpha) - s\,Z_s(s{+}1,\alpha)\Big]_{s=0} \nonumber\\
  &= -Z(1,\alpha). \label{eq:bridge}
\end{align}
The second term vanishes because of the prefactor $s|_{s=0}=0$. Therefore
\begin{equation}\label{eq:DW_route1_final}
  DW^{(1)} = -x^2\,Z(1,\alpha).
\end{equation}

\paragraph{Differentiation $\Rightarrow$ regularization}

Applying $D=R\partial_R$ to Eq.~\eqref{eq:wbare} and differentiating term by term
(a step that is not a priori justified) we obtain
\begin{align}
    DW_{\rm bare}=\sum_{n=1}^{\infty}\left[n^2-\frac{n^2x^2}{n^2+\alpha}\right].
\end{align}
Both sums are divergent. We regularize $\sum n^2\to Z(0, \alpha) = 0$, and
$\sum n^2/(n^2+\alpha) = Z(1, \alpha)$. Recall that since $Z(s,\alpha)=H(s{-}1,\alpha)-\alpha H(s,\alpha)$
and $H(s,\alpha)$ has poles only at $s=\tfrac12,-\tfrac12,\dots$\,,
the continuation is regular at $s=1$ and gives
$Z(1,\alpha)=H(0,\alpha)-\alpha H(1,\alpha)$,
where $H(1,\alpha)=\sum(n^2+\alpha)^{-1}$ converges absolutely.
Thus
\begin{align}\label{eq:DW_route2}
    DW^{(2)}=-x^2\,Z(1,\alpha).
\end{align}
This route is justified a posteriori: the bridge identity~\eqref{eq:bridge}
and $Z(0,\alpha)=0$ guarantee that it agrees with Route~1.

\subsection*{Some observations}
The two routes agree: $DW^{(1)}=DW^{(2)}=-x^2 Z(1,\alpha)$, as established by the bridge identity~\eqref{eq:bridge}. The agreement relies on the fact that $Z(0, \alpha) = 0$ for all $\alpha$. Had this not held, the extra terms -- $2x^2\log R \partial_\alpha Z(0,\alpha)$ in Route~1 and $\sum n^2$ in Route~2 respectively-- would have spoiled the equality. In the counterexample $\sum(n+a)$ mentioned above, the analogous quantity $\zeta_H(0,a)=\tfrac12-a$ is \emph{not} $a$-independent, and the extra term does not vanish---this is the source of the discrepancy.

\subsection*{Evaluation of $Z(1,\alpha)$}

From~\eqref{eq:Zdecomp} at $s=1$:
\begin{equation}\label{eq:Z1decomp}
    Z(1,\alpha) = H(0,\alpha)-\alpha H(1,\alpha).
\end{equation}

\paragraph{$H(0,\alpha)$.}
The asymptotic expansion $(n^2+\alpha)^{-s}=n^{-2s}[1-s\alpha n^{-2}+\cdots]$ gives $H(s,\alpha)=\zeta(2s)-s\alpha\,\zeta(2s{+}2)+\cdots$, all $\alpha$-dependent terms carry a factor of~$s$ and vanish at $s=0$. Hence
\begin{equation}
    H(0,\alpha)=\zeta(0)=-\tfrac{1}{2},
\end{equation}
independently of~$\alpha$.

\paragraph{$H(1,\alpha)$.} For $\alpha\notin\{-n^2\}_{n\geq1}$, the series $H(1,\alpha)=\sum_{n=1}^\infty (n^2+\alpha)^{-1}$ converges absolutely. We use the Mittag--Leffler partial-fraction expansion of $\pi\cot(\pi w)$. Starting from
\begin{equation}
  \pi\cot(\pi w) = \frac{1}{w}+2w\sum_{n=1}^\infty\frac{1}{w^2-n^2},
\end{equation}
substitute $w=i\sqrt{\alpha}$ (with $\cot(ix)=-i\coth(x)$) to obtain
\begin{equation}\label{eq:ML}
  \pi\sqrt{\alpha}\,\coth\!\big(\pi\sqrt{\alpha}\,\big)
  = 1+2\alpha\sum_{n=1}^\infty\frac{1}{n^2+\alpha}.
\end{equation}
Therefore
\begin{equation}\label{eq:H1}
  H(1,\alpha) = \frac{1}{2\alpha}\!\left[\pi\sqrt{\alpha}\,\coth\!\big(\pi\sqrt{\alpha}\,\big)-1\right]
  = \frac{\pi\coth(\pi\sqrt{\alpha}\,)}{2\sqrt{\alpha}} - \frac{1}{2\alpha}.
\end{equation}
Substituting into \eqref{eq:Z1decomp}:
\begin{align}
  Z(1,\alpha) &= -\tfrac12 - \alpha\!\left[\frac{\pi\coth(\pi\sqrt{\alpha}\,)}{2\sqrt{\alpha}}-\frac{1}{2\alpha}\right]\nonumber\\
  &= -\tfrac12 - \frac{\pi\sqrt{\alpha}}{2}\coth\!\big(\pi\sqrt{\alpha}\,\big)+\tfrac12\nonumber\\
  &= -\frac{\pi\sqrt{\alpha}}{2}\,\coth\!\big(\pi\sqrt{\alpha}\,\big).
\end{align}
With $\nu\equiv\sqrt{\alpha}=\sqrt{x^2-\tfrac14}$:
\begin{equation}
  Z(1,\alpha) = -\frac{\pi\nu}{2}\,\coth(\pi\nu).
\end{equation}
From $DW=-x^2 Z(1,\alpha)$:
\begin{equation}\label{eq:DWfinal}
  DW(x) = \frac{\pi x^2}{2}\,\nu(x)\,\coth\!\big(\pi\nu(x)\big),
  \qquad \nu=\sqrt{x^2-\tfrac14}.
\end{equation}
This is identical to the result stated in Eq. \eqref{eq:DW_closed}.

\bibliography{refs}

\begin{thebibliography}{15}%
\makeatletter
\providecommand \@ifxundefined [1]{%
 \@ifx{#1\undefined}
}%
\providecommand \@ifnum [1]{%
 \ifnum #1\expandafter \@firstoftwo
 \else \expandafter \@secondoftwo
 \fi
}%
\providecommand \@ifx [1]{%
 \ifx #1\expandafter \@firstoftwo
 \else \expandafter \@secondoftwo
 \fi
}%
\providecommand \natexlab [1]{#1}%
\providecommand \enquote  [1]{``#1''}%
\providecommand \bibnamefont  [1]{#1}%
\providecommand \bibfnamefont [1]{#1}%
\providecommand \citenamefont [1]{#1}%
\providecommand \href@noop [0]{\@secondoftwo}%
\providecommand \href [0]{\begingroup \@sanitize@url \@href}%
\providecommand \@href[1]{\@@startlink{#1}\@@href}%
\providecommand \@@href[1]{\endgroup#1\@@endlink}%
\providecommand \@sanitize@url [0]{\catcode `\\12\catcode `\$12\catcode
  `\&12\catcode `\#12\catcode `\^12\catcode `\_12\catcode `\%12\relax}%
\providecommand \@@startlink[1]{}%
\providecommand \@@endlink[0]{}%
\providecommand \url  [0]{\begingroup\@sanitize@url \@url }%
\providecommand \@url [1]{\endgroup\@href {#1}{\urlprefix }}%
\providecommand \urlprefix  [0]{URL }%
\providecommand \Eprint [0]{\href }%
\providecommand \doibase [0]{https://doi.org/}%
\providecommand \selectlanguage [0]{\@gobble}%
\providecommand \bibinfo  [0]{\@secondoftwo}%
\providecommand \bibfield  [0]{\@secondoftwo}%
\providecommand \translation [1]{[#1]}%
\providecommand \BibitemOpen [0]{}%
\providecommand \bibitemStop [0]{}%
\providecommand \bibitemNoStop [0]{.\EOS\space}%
\providecommand \EOS [0]{\spacefactor3000\relax}%
\providecommand \BibitemShut  [1]{\csname bibitem#1\endcsname}%
\let\auto@bib@innerbib\@empty
\bibitem [{\citenamefont {Jafferis}\ \emph {et~al.}(2011)\citenamefont
  {Jafferis}, \citenamefont {Klebanov}, \citenamefont {Pufu},\ and\
  \citenamefont {Safdi}}]{Jafferis:2011}%
  \BibitemOpen
  \bibfield  {author} {\bibinfo {author} {\bibfnamefont {D.~L.}\ \bibnamefont
  {Jafferis}}, \bibinfo {author} {\bibfnamefont {I.~R.}\ \bibnamefont
  {Klebanov}}, \bibinfo {author} {\bibfnamefont {S.~S.}\ \bibnamefont {Pufu}},\
  and\ \bibinfo {author} {\bibfnamefont {B.~R.}\ \bibnamefont {Safdi}},\
  }\bibfield  {title} {\bibinfo {title} {Towards the {$F$}-theorem:
  {$\mathcal{N}{=}2$} field theories on the three-sphere},\ }\href
  {https://doi.org/10.1007/JHEP06(2011)102} {\bibfield  {journal} {\bibinfo
  {journal} {JHEP}\ }\textbf {\bibinfo {volume} {06}},\ \bibinfo {pages}
  {102}},\ \Eprint {https://arxiv.org/abs/1103.1181} {arXiv:1103.1181 [hep-th]}
  \BibitemShut {NoStop}%
\bibitem [{\citenamefont {Klebanov}\ \emph {et~al.}(2011)\citenamefont
  {Klebanov}, \citenamefont {Pufu},\ and\ \citenamefont
  {Safdi}}]{Klebanov:2011}%
  \BibitemOpen
  \bibfield  {author} {\bibinfo {author} {\bibfnamefont {I.~R.}\ \bibnamefont
  {Klebanov}}, \bibinfo {author} {\bibfnamefont {S.~S.}\ \bibnamefont {Pufu}},\
  and\ \bibinfo {author} {\bibfnamefont {B.~R.}\ \bibnamefont {Safdi}},\
  }\bibfield  {title} {\bibinfo {title} {{$F$}-theorem without supersymmetry},\
  }\href {https://doi.org/10.1007/JHEP10(2011)038} {\bibfield  {journal}
  {\bibinfo  {journal} {JHEP}\ }\textbf {\bibinfo {volume} {10}},\ \bibinfo
  {pages} {038}},\ \Eprint {https://arxiv.org/abs/1105.4598} {arXiv:1105.4598
  [hep-th]} \BibitemShut {NoStop}%
\bibitem [{\citenamefont {Casini}\ and\ \citenamefont
  {Huerta}(2012)}]{CasiniHuerta:2012}%
  \BibitemOpen
  \bibfield  {author} {\bibinfo {author} {\bibfnamefont {H.}~\bibnamefont
  {Casini}}\ and\ \bibinfo {author} {\bibfnamefont {M.}~\bibnamefont
  {Huerta}},\ }\bibfield  {title} {\bibinfo {title} {On the rg running of the
  entanglement entropy of a circle},\ }\href
  {https://doi.org/10.1103/PhysRevD.85.125016} {\bibfield  {journal} {\bibinfo
  {journal} {Phys. Rev. D}\ }\textbf {\bibinfo {volume} {85}},\ \bibinfo
  {pages} {125016} (\bibinfo {year} {2012})},\ \Eprint
  {https://arxiv.org/abs/1202.5650} {arXiv:1202.5650 [hep-th]} \BibitemShut
  {NoStop}%
\bibitem [{\citenamefont {Zamolodchikov}(1986)}]{Zamolodchikov:1986}%
  \BibitemOpen
  \bibfield  {author} {\bibinfo {author} {\bibfnamefont {A.~B.}\ \bibnamefont
  {Zamolodchikov}},\ }\bibfield  {title} {\bibinfo {title} {``irreversibility''
  of the flux of the renormalization group in a 2d field theory},\ }\href@noop
  {} {\bibfield  {journal} {\bibinfo  {journal} {JETP Lett.}\ }\textbf
  {\bibinfo {volume} {43}},\ \bibinfo {pages} {730} (\bibinfo {year} {1986})},\
  \bibinfo {note} {english translation of Pis'ma Zh. Eksp. Teor. Fiz. 43 (1986)
  565--567}\BibitemShut {NoStop}%
\bibitem [{\citenamefont {Affleck}(1986)}]{Affleck:1986}%
  \BibitemOpen
  \bibfield  {author} {\bibinfo {author} {\bibfnamefont {I.}~\bibnamefont
  {Affleck}},\ }\bibfield  {title} {\bibinfo {title} {Universal term in the
  free energy at a critical point and the conformal anomaly},\ }\href
  {https://doi.org/10.1103/PhysRevLett.56.746} {\bibfield  {journal} {\bibinfo
  {journal} {Phys. Rev. Lett.}\ }\textbf {\bibinfo {volume} {56}},\ \bibinfo
  {pages} {746} (\bibinfo {year} {1986})}\BibitemShut {NoStop}%
\bibitem [{\citenamefont {Bl{\"o}te}\ \emph {et~al.}(1986)\citenamefont
  {Bl{\"o}te}, \citenamefont {Cardy},\ and\ \citenamefont
  {Nightingale}}]{BloteCardyNightingale:1986}%
  \BibitemOpen
  \bibfield  {author} {\bibinfo {author} {\bibfnamefont {H.~W.~J.}\
  \bibnamefont {Bl{\"o}te}}, \bibinfo {author} {\bibfnamefont {J.~L.}\
  \bibnamefont {Cardy}},\ and\ \bibinfo {author} {\bibfnamefont {M.~P.}\
  \bibnamefont {Nightingale}},\ }\bibfield  {title} {\bibinfo {title}
  {Conformal invariance, the central charge, and universal finite-size
  amplitudes at criticality},\ }\href
  {https://doi.org/10.1103/PhysRevLett.56.742} {\bibfield  {journal} {\bibinfo
  {journal} {Phys. Rev. Lett.}\ }\textbf {\bibinfo {volume} {56}},\ \bibinfo
  {pages} {742} (\bibinfo {year} {1986})}\BibitemShut {NoStop}%
\bibitem [{\citenamefont {Nakayama}(2015)}]{Nakayama:2013is}%
  \BibitemOpen
  \bibfield  {author} {\bibinfo {author} {\bibfnamefont {Y.}~\bibnamefont
  {Nakayama}},\ }\bibfield  {title} {\bibinfo {title} {{Scale invariance vs
  conformal invariance}},\ }\href
  {https://doi.org/10.1016/j.physrep.2014.12.003} {\bibfield  {journal}
  {\bibinfo  {journal} {Phys. Rept.}\ }\textbf {\bibinfo {volume} {569}},\
  \bibinfo {pages} {1} (\bibinfo {year} {2015})},\ \Eprint
  {https://arxiv.org/abs/1302.0884} {arXiv:1302.0884 [hep-th]} \BibitemShut
  {NoStop}%
\bibitem [{\citenamefont {Birrell}\ and\ \citenamefont
  {Davies}(1982)}]{BirrellDavies:1982}%
  \BibitemOpen
  \bibfield  {author} {\bibinfo {author} {\bibfnamefont {N.~D.}\ \bibnamefont
  {Birrell}}\ and\ \bibinfo {author} {\bibfnamefont {P.~C.~W.}\ \bibnamefont
  {Davies}},\ }\href {https://doi.org/10.1017/CBO9780511622632} {\emph
  {\bibinfo {title} {Quantum Fields in Curved Space}}},\ Cambridge Monographs
  on Mathematical Physics\ (\bibinfo  {publisher} {Cambridge University
  Press},\ \bibinfo {address} {Cambridge},\ \bibinfo {year} {1982})\BibitemShut
  {NoStop}%
\bibitem [{\citenamefont {Vassilevich}(2003)}]{Vassilevich:2003}%
  \BibitemOpen
  \bibfield  {author} {\bibinfo {author} {\bibfnamefont {D.~V.}\ \bibnamefont
  {Vassilevich}},\ }\bibfield  {title} {\bibinfo {title} {Heat kernel
  expansion: user's manual},\ }\href
  {https://doi.org/10.1016/j.physrep.2003.09.002} {\bibfield  {journal}
  {\bibinfo  {journal} {Phys. Rept.}\ }\textbf {\bibinfo {volume} {388}},\
  \bibinfo {pages} {279} (\bibinfo {year} {2003})},\ \Eprint
  {https://arxiv.org/abs/hep-th/0306138} {arXiv:hep-th/0306138} \BibitemShut
  {NoStop}%
\bibitem [{\citenamefont {Pufu}(2017)}]{Pufu:2016}%
  \BibitemOpen
  \bibfield  {author} {\bibinfo {author} {\bibfnamefont {S.~S.}\ \bibnamefont
  {Pufu}},\ }\bibfield  {title} {\bibinfo {title} {The f-theorem and
  f-maximization},\ }\href {https://doi.org/10.1088/1751-8121/aa6765}
  {\bibfield  {journal} {\bibinfo  {journal} {Journal of Physics A:
  Mathematical and Theoretical}\ }\textbf {\bibinfo {volume} {50}},\ \bibinfo
  {pages} {443008} (\bibinfo {year} {2017})},\ \Eprint
  {https://arxiv.org/abs/1608.02960} {arXiv:1608.02960 [hep-th]} \BibitemShut
  {NoStop}%
\bibitem [{\citenamefont {Abramowitz}\ and\ \citenamefont
  {Stegun}(1965)}]{AbramowitzStegun:1965}%
  \BibitemOpen
  \bibinfo {editor} {\bibfnamefont {M.}~\bibnamefont {Abramowitz}}\ and\
  \bibinfo {editor} {\bibfnamefont {I.~A.}\ \bibnamefont {Stegun}},\ eds.,\
  \href@noop {} {\emph {\bibinfo {title} {Handbook of Mathematical Functions:
  with Formulas, Graphs, and Mathematical Tables}}},\ \bibinfo {series}
  {National Bureau of Standards Applied Mathematics Series}, Vol.~\bibinfo
  {volume} {55}\ (\bibinfo  {publisher} {U.S. Government Printing Office},\
  \bibinfo {year} {1965})\BibitemShut {NoStop}%
\bibitem [{\citenamefont {Gradshteyn}\ and\ \citenamefont
  {Ryzhik}(2007)}]{GradshteynRyzhik:2007}%
  \BibitemOpen
  \bibfield  {author} {\bibinfo {author} {\bibfnamefont {I.~S.}\ \bibnamefont
  {Gradshteyn}}\ and\ \bibinfo {author} {\bibfnamefont {I.~M.}\ \bibnamefont
  {Ryzhik}},\ }\href@noop {} {\emph {\bibinfo {title} {Table of Integrals,
  Series, and Products}}},\ \bibinfo {edition} {7th}\ ed.,\ edited by\ \bibinfo
  {editor} {\bibfnamefont {A.}~\bibnamefont {Jeffrey}}\ and\ \bibinfo {editor}
  {\bibfnamefont {D.}~\bibnamefont {Zwillinger}}\ (\bibinfo  {publisher}
  {Academic Press},\ \bibinfo {year} {2007})\BibitemShut {NoStop}%
\bibitem [{\citenamefont {Komargodski}\ and\ \citenamefont
  {Schwimmer}(2011)}]{KomargodskiSchwimmer:2011}%
  \BibitemOpen
  \bibfield  {author} {\bibinfo {author} {\bibfnamefont {Z.}~\bibnamefont
  {Komargodski}}\ and\ \bibinfo {author} {\bibfnamefont {A.}~\bibnamefont
  {Schwimmer}},\ }\bibfield  {title} {\bibinfo {title} {On renormalization
  group flows in four dimensions},\ }\href
  {https://doi.org/10.1007/JHEP12(2011)099} {\bibfield  {journal} {\bibinfo
  {journal} {JHEP}\ }\textbf {\bibinfo {volume} {12}},\ \bibinfo {pages}
  {099}},\ \Eprint {https://arxiv.org/abs/1107.3987} {arXiv:1107.3987 [hep-th]}
  \BibitemShut {NoStop}%
\bibitem [{\citenamefont {Santoni}\ and\ \citenamefont
  {Scardino}()}]{companion}%
  \BibitemOpen
  \bibfield  {author} {\bibinfo {author} {\bibfnamefont {G.}~\bibnamefont
  {Santoni}}\ and\ \bibinfo {author} {\bibfnamefont {F.}~\bibnamefont
  {Scardino}},\ }\bibfield  {title} {\bibinfo {title} {(to appear)},\
  }\href@noop {} {\ }\BibitemShut {NoStop}%
\bibitem [{\citenamefont {Kirsten}(2001)}]{Kirsten:2001wz}%
  \BibitemOpen
  \bibfield  {author} {\bibinfo {author} {\bibfnamefont {K.}~\bibnamefont
  {Kirsten}},\ }\href@noop {} {\emph {\bibinfo {title} {{Spectral functions in
  mathematics and physics}}}}\ (\bibinfo {year} {2001})\BibitemShut {NoStop}%
\end{thebibliography}%
\end{document}